\documentclass[aps,twocolumn,epsf,floats,prl]{revtex4}
\usepackage{graphics,graphicx,epsfig}
\usepackage{amssymb}
\usepackage{epsf,epstopdf,wrapfig}

\begin{document}

\title{Toward  a statistical mechanics of four letter words}

\author{Greg J Stephens$^a$ and William Bialek$^b$}

\affiliation{$^{a,b}$Joseph Henry Laboratories of Physics,
$^{a,b}$Lewis--Sigler Institute for Integrative Genomics, 
$^a$Center for the Study of Mind, Brain and Behavior, and 
$^b$Princeton Center for Theoretical Physics, Princeton University,
Princeton, New Jersey 08544 USA}

\date{\today}

\begin{abstract} 
We consider words as a network of interacting letters, and approximate the probability distribution of states taken on by this network.  Despite the intuition that the rules of English spelling are highly combinatorial (and arbitrary), we find that maximum entropy models consistent with pairwise correlations among letters provide a surprisingly good approximation to the full statistics of four letter words, capturing $\sim 92\%$ of the multi--information among letters and even `discovering' real words that were not represented in the data from which the pairwise correlations were estimated.  The maximum entropy model defines an energy landscape on the space of possible words, and local minima in this landscape account for nearly two--thirds of words used in written English.
\end{abstract}

\maketitle

Many complex systems convey an impression of order that is not so easily captured by the traditional tools of theoretical physics.  Thus, it is not clear what sort of order parameter or correlation function we should compute to detect that natural images are composed of solid objects \cite{images}, nor is it obvious what features of the amino acid sequence distinguish foldable  proteins from random polymers. Recently, several groups have tried to simplify the problem of characterizing order in biological systems using the classical idea of maximum entropy \cite{jaynes_57}.  Maximum entropy models consistent with pairwise correlations among neurons have proven surprisingly effective in describing the patterns of activity in real networks ranging from the retina \cite{schneidman+al_06,shlens+al_06} to the cortex \cite{more_neurons2}; these models are identical to the Ising models of statistical mechanics, which have long been explored as abstract models for neural networks \cite{hopfield_82}.  
Similar methods have been used to analyze biochemical \cite{tkacik_07a} and genetic \cite{psu_group} networks, and these approaches are connected to an independent stream of work arguing that pairwise correlations among amino acids may be sufficient to define functional proteins \cite{bialek+ranganathan_07}.  Because of the immediate connection to statistical mechanics, this work also provides a natural path for extrapolating to the collective behavior of large networks, starting with real data \cite{tkacik+al_06}.
Here we test the limits of these ideas   constructing maximum entropy models for the sequence of letters in words.

As non--native speakers know well, the rules of English spelling seem arbitrary and almost paradigmatically combinatorial (i before e except after c).  In contrast, the whole point of maximum entropy constructions based on pairwise correlations is to ignore such higher order, combinatorial effects \cite{schneidman+al_03}.  We thus suspect that the statistics of letters in words will provide an interesting test case.  There is  a long history of statistical approaches in the analysis of language, including applications of maximum entropy ideas \cite{berger+al_96}, while  opposition to such statistical approaches was at the foundation of forty years of linguistic theory \cite{chomsky_56,abney_96}; for a recent view of these debates see Ref \cite{pereira_00}.  Our goal here is not to enter into these controversies about language in the broad sense, but rather to test the power of pairwise interactions to capture seemingly complex structure.

\begin{table}[b]
\begin{center}
\begin{tabular}{| c | c | c | c |c |} \hline
& \multicolumn{2}{|c|}{Jane Austen} & \multicolumn{2}{|c|}{ANC} \\ \hline
rank & word &  probability & word &probability  \\ [0.5ex] \hline
1 & that   & 0.0649   & with & 0.0614 \\
\hline
2 &with &  0.0468   & that & 0.0600 \\
\hline
3 &have &  0.0424   &  from & 0.0368 \\
\hline
4 &very & 0.0289  & this & 0.0366  \\
\hline
5& been & 0.0243 & were& 0.0279 \\    
\hline
6 &were & 0.0236   &  they & 0.0267\\
\hline
7& they & 0.0230  &  have & 0.0234\\ 
\hline
8 & from & 0.0225   & said & 0.0176 \\ 
\hline
9& this & 0.0199   & when &0.0164 \\
\hline
10 &what & 0.0190 & will & 0.0158 \\
\hline
 20 & than & 0.0149 & more & 0.0109\\
 \hline
 50 & over & 0.0045 & cell & 0.0035\\
 \hline
 100  & word &   0.0021 & area & 0.0017\\
 \hline
last & hazy & $7.38\times10^{-6}$&warn&  $3.90\times10^{-5}$   \\
 \hline
\end{tabular}
\end{center}
\caption{Four letter words in two large corpora, the novels of Jane Austen \cite{austen} and the American National Corpus \cite{anc}.}
\label{words}
\end{table}

Even with only four letters, there are $N=(26)^4 = 456,976$ possible words, but  only a tiny fraction of these are real (or even `legal') words in English.  Our problem thus is easy to state:   are maximum entropy models powerful enough to capture this restriction of vocabulary?

\begin{figure*}
\includegraphics[width=6in]{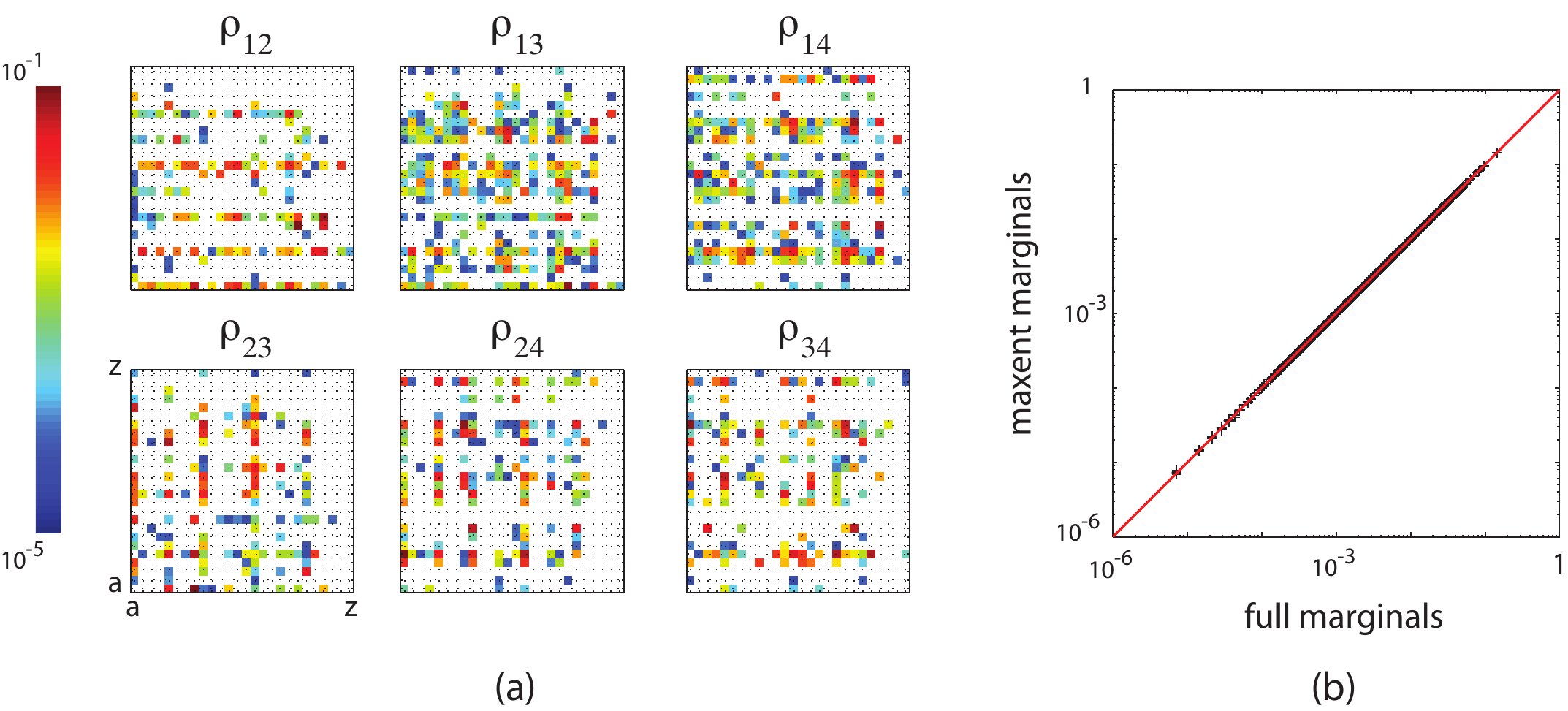}
\caption{(a) The six pairwise marginal distributions of four-letter words sampled from the Jane Austen corpus.  Common letter pairs such as ``th"  in $\rho_{12}$ are apparent in their large marginal probability.  (b) The iterative scaling algorithm solves the constrained maximization problem to high precision.   All pairwise marginal components of the full distribution compared to the marginals constructed from the computed maximum entropy distribution}.  
\label{fig:margRecon}
\end{figure*}

To analyze the interactions among letters, we use two large corpora: a collection of novels from Jane Austen \cite{austen}, and a large sampling of American English contained in the American National Corpus (ANC) \cite{anc}.  To control for potential typographic errors, words were also checked against a large dictionary
database \cite{dictionary}.  Out of 676,302 total words in our Austen corpus there were 7114 unique words, 763 of which 
were four letter words;  the four letter words occurred in the corpus a total of 135,441 times.   We used the second release of the ANC  
($\sim 2\times 10^7$ words), and restricted ourselves to words used more than 100 times,  providing 798 unique  four letter words occurring  2,179,108 times.   These numbers indicate that we can sample the distribution of four letter words with reasonable confidence.  The most common words and their probabilities  are shown in Table \ref{words}.

We are interested in the full joint distribution $P(\ell_1 , \ell_2 , \ell_3 , \ell_4)$ of letters in a four letter word.  The maximum possible entropy of this distribution is  $S_{\rm rand} =4  \log_2 (26) = 18.802 \,{\rm bits}$.  Letters occur with different probabilities, however, so even if we compose words by choosing letters independently and at random out of real text, the entropy will be lower than this.  A more precisely defined `independent model' is the approximation
\begin{equation}
P(\ell_1 , \ell_2 , \ell_3 , \ell_4) \approx P^{(1)} = \prod_{{\rm i}=1}^4 P(\ell_{\rm i}) ,
\end{equation}
where we note that each of the $P(\ell_{\rm i})$ is different because letters are used differently at different positions in the word.  In the Austen corpus, this independent model has an entropy $S_{\rm ind} = 14.083\pm 0.001\,{\rm bits}$, while the full distribution has entropy of just $S_{\rm full} = 6.92\pm 0.003\,{\rm bits}$ \cite{technical}.  The difference between these quantities is the multi--information,
$I \equiv S_{\rm ind} - S_{\rm full} = 7.163\,{\rm bits}$, which measures the amount of structure or correlation in the joint distribution of letters that form real words.  Thus,
spelling rules restrict the vocabulary by a factor of $2^I \sim 143$ relative to the number of words that would be allowed if letters were chosen independently.

Maximum entropy models based on pairwise correlations are equivalent to Boltzmann distributions with pairwise interactions among the elements of the system (see, for example Ref \cite{schneidman+al_03}); in our case this means approximating $P(\ell_1 , \ell_2 , \ell_3 ,\ell_4 ) \approx P^{(2)}$,
\begin{eqnarray}
P^{(2)} (\ell_1 , \ell_2 , \ell_3 , \ell_4 ) = {1\over Z} \exp\left[
-\sum_{{\rm i} > {\rm j}} V_{\rm ij} (\ell_{\rm i} , \ell_{\rm j})
\right],
\label{boltz}
\end{eqnarray}
where the $V_{\rm ij}$ are `interaction potentials' between pairs of letters and  $Z$ serves to normalize the distribution;  because the order of letters matters, there are six independent potentials.  Each potential is a $26\times 26$ matrix, but the zero of energy is arbitrary, so this model has  $6\times (26^2 -1) = 4050$ parameters, more than $100\times$ less than the number of possible states.
Note that  interactions extend across the full length of the word, so that the maximum entropy model built from pairwise correlations is very different from a Markov model which only allows each letter to interact with its neighbor. 

We determine the interaction potentials $V_{\rm ij}$
by matching to the pairwise marginal distributions, that is by solving the six coupled sets of $26^2$ equations:
\begin{equation}
\rho_{12}(\ell , \ell') = \sum_{\ell_3 , \ell_4} P^{(2)} (\ell , \ell' , \ell_3 , \ell_4 ),
\end{equation}
and similarly for the other five pairs.
As shown in Fig \ref{fig:margRecon}a, the pairwise marginals sampled from English are highly structured; many entries in the marginal distributions are exactly zero, even in corpora with millions of words.

Construction of maximum entropy models for large systems is difficult  \cite{broderick+al_07}, but for $\sim 5\times 10^5$ states as in our problem relatively simple algorithms suffice \cite{darroch+ratcliff_72}.  We see from Fig \ref{fig:margRecon}b that these methods succeed in matching the observed pairwise marginals with high precision.

\begin{figure}[b]
\includegraphics[width=0.90\columnwidth,keepaspectratio=true]{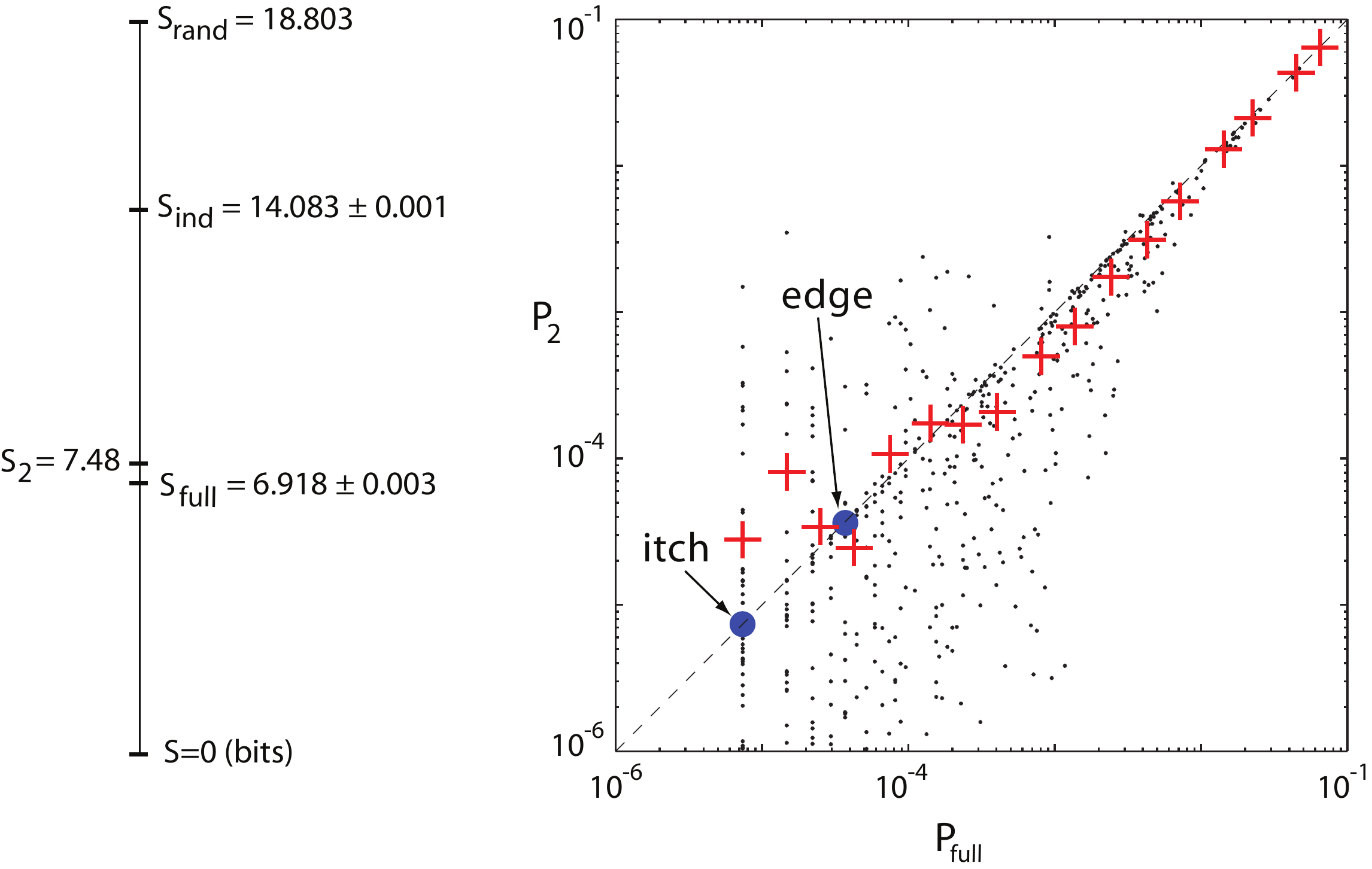}
\caption{(left) The pairwise maximum entropy model provides an excellent approximation to the full distribution of
four-letter words, capturing $92\%$ of the multi-information.
(right-dots) Scatter plot of the four letter word probabilities in the full distribution $P_{sampled}$ vs.~the corresponding probabilities in the maximum
entropy distribution $P_2$.  (right-red crosses) To facilitate the comparison we divided the full probability into 20 equally log-spaced
bins and computed the mean maximum entropy probability conditioned on the states in the full distribution within each bin.  The dashed line marks
the identity.  (right-blue circles) Even for small probabilities, there are still words such as `edge' and `itch'  whose states are well-captured by the pairwise model.}
\label{fig:maxS}
\end{figure}

As shown at left in Figure \ref{fig:maxS}, the  maximum entropy model with pairwise interactions does a surprisingly good job in capturing the structure of the full distribution.  In the Austen corpus the model predicts an entropy $S_2 = 7.48$ bits, which means that it captures $92\%$ of the multi--information, and similar results are found with the ANC, where we capture $89\%$ of the multi--information.  
Pairwise interactions thus restrict the vocabulary by a factor of $2^{S_{\rm ind} - S_2} \sim 100$ relative to the words which are possible by choosing letters independently.

While very good, the maximum entropy model is of course not perfect, as we can see at right in Fig \ref{fig:maxS}.  Here we see that the probabilities of the individual words predicted by the model agree only approximately with the observed probabilities.  There is good agreement on average, especially for the more common words, but substantial scatter.  On the other hand, there are particular words with low probability, whose frequency of use is predicted with high accuracy. We have singled out two of these, `edge' and `itch,' as somewhat surprising.  Each contains sounds composed of three consonants in sequence, and we might have expected that to predict the frequency of these combinations one would need to incorporate three--body interactions among the letters, but it seems that pairwise interactions are sufficient.

\begin{figure}
\begin{center}
\includegraphics[width=0.95\columnwidth]{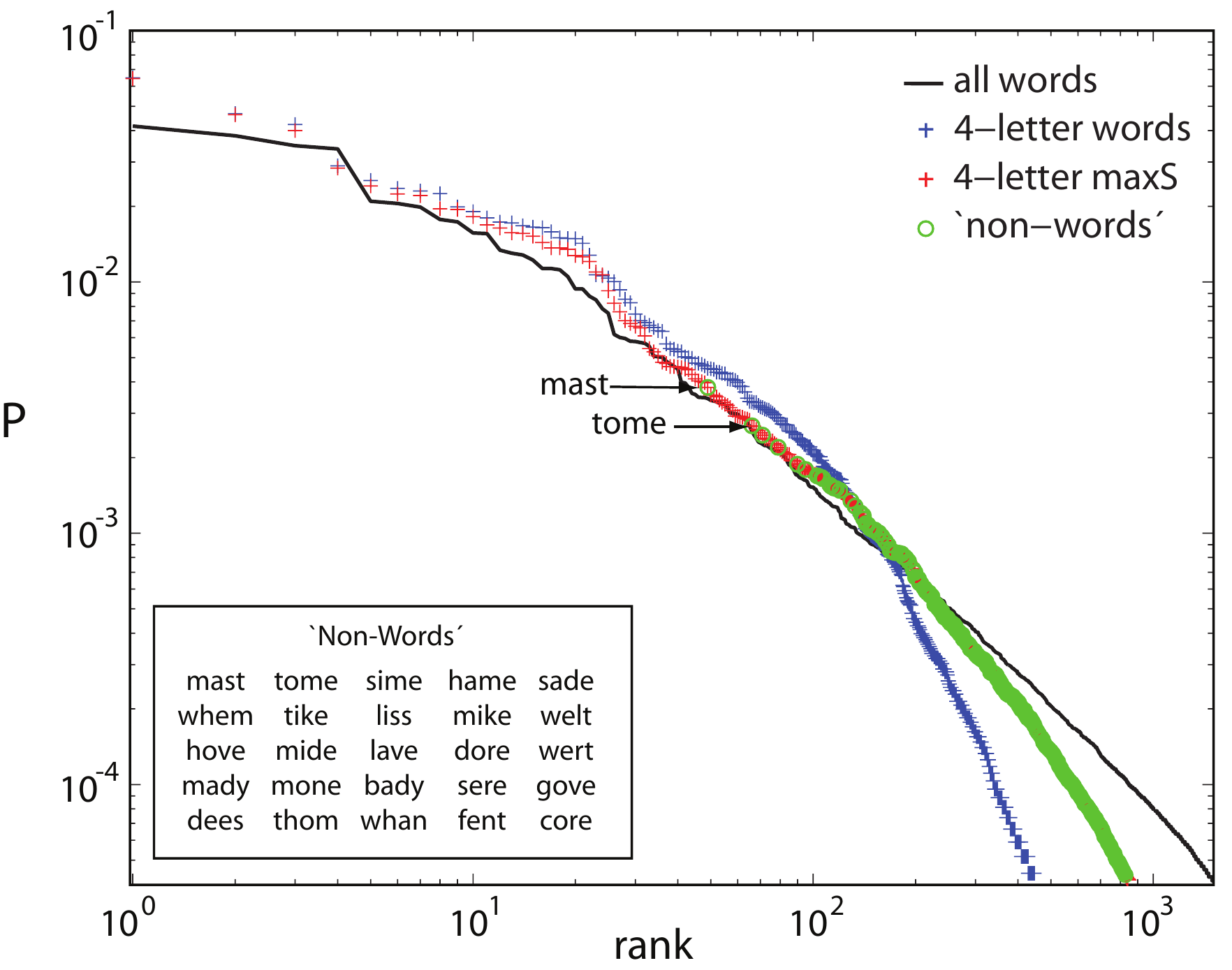}
\end{center}
\caption{The Zipf plot for all words in the corpus (black line),  four letter words in the corpus (blue crosses), and four letter words in
the maximum entropy model (red crosses).  Green circles denote `non-words', states in the maximum entropy model that didn't appear in the corpus.  The 25 most likely `non--words' are shown in the text
inset (ordered in decreasing probability from left to right and top to bottom).  Some of these are recognizable as real words that just did not appear in the corpus, and even the others have plausible 
spelling.}
\label{fig:zipf}
\end{figure}

Another way of looking at structure in the distribution of words is the Zipf plot \cite{zipf_32}, the probability of a word's use as a function of its rank  (Fig \ref{fig:zipf}).  If we look at all words, we recover the approximate power law which initially intrigued Zipf; when we confine our attention to four letter words, the long tail is cut off \cite{li_92}.  The maximum entropy model does a good job of reproducing the observed Zipf plot, but removes some weight from the bulk of the distribution and reassigns it to words which do not occur in the corpus, repopulating the tail.  Importantly, as shown in the inset to Fig \ref{fig:zipf}, many of these `non--words' are perfectly good English words which  happen not to have been used by Jane Austen.  Quantitatively, the maximum entropy models assigns $15\%$ of the probability to words which do not appear in the corpus, but of these $1/5$ are words that can be found in the dictionary, and the same factor for the `correct discovery' of new words is found with the ANC.  Although somewhat subjective, we note that even words not found in the dictionary seem to be speakable combinations of letters, not obviously violating known spelling rules.

If we take Equation (\ref{boltz}) seriously as a statistical mechanics problem, then we have a constructed an energy landscape on the space of words, much in the spirit of Hopfield's energy landscape for the states of neural networks \cite{hopfield_82}.  In this landscape there are local minima, combinations of letters for which any single letter change will result in an increase in the energy.  In the maximum entropy model for the ANC, there are 136 of these local minima, of which 118 are real English words, capturing nearly two thirds ($63.5\%$) of the full distribution; similar results are obtained in the Austen corpus.  We note that such `stable words' have the property that any single letter spelling error can always be corrected by relaxing to the nearest local minimum of the energy.  It is tempting to suggest that if we could construct the energy landscape for sentences (rather than for words), then almost all legal sentences would be locally stable.

To summarize, in English we use only a very small fraction ($\sim 1/3700$) of the roughly half million possible four letter combinations.  
The hierarchy of maximum entropy constructions \cite{schneidman+al_03} allows us to decompose these spelling rules into contributions from interactions of different order.  A significant factor ($\sim 1/26$) comes from the unequal probabilities with which individual letters are used,  a larger factor ($\sim 1/100$) comes from the pairwise interactions among letters, and higher order interactions contribute only a small factor ($\sim 1/1.5$).   The pairwise model represents an enormous simplification, which nonetheless has the power to capture most of the structure in the distribution of letters and even to discover combinations of letters that are legal but unused in the corpora from which we have learned.  The analogy to statistical mechanics also invites us to think about the way in which combinations of competing interactions enforce a complex landscape, singling out words which can be transmitted stably even in the presence of errors.  Although our primary interest has been to test the power of the maximum entropy models, these ideas of generalization and error correction seem relevant to understanding the cognitive processing of text \cite{pereira_00,dehaene+al_05}.

\begin{acknowledgments}
We thank D Chigirev, SE Palmer, E Schneidman \& G Tka\v{c}ik for helpful discussions.  This work was supported in part by National Science Foundation Grants IIS--0613435 and PHY--0650617,
by National Institutes of Health Grants P50 GM071508 and T32 MH065214, and by  the Swartz Foundation.  
\end{acknowledgments}


\begin{thebibliography}{99}
\bibitem{images}
See, for example,  D Mumford, 
Empirical statistics and stochastic models for visual signals, 
in {\em New Directions in Statistical Signal Processing:  From Systems to Brain,} S Haykin et al, eds, pp 3--34 (MIT Press, Cambridge, 2005).
\bibitem{jaynes_57}
ET Jaynes,
Information theory and statistical mechanics. 
{\em Phys Rev} {\bf 106,} 62--79 (1957).
\bibitem{schneidman+al_06}
E Schneidman, MJ Berry II, R Segev \& W Bialek, 
Weak pairwise correlations imply strongly correlated network states in a neural population.  
{\em Nature} {\bf 440,} 1007--1012 (2006); q--bio.NC/0512013.
\bibitem{shlens+al_06}
J Shlens et al, 
The structure of multi--neuron firing patterns in primate retina.
{\em J Neurosci} {\bf 26,} 8254--8266 (2006).
\bibitem{more_neurons2}
See the presentations at the  Society for Neuroscience, {\tt http://www.sfn.org/am2007/}:
IE Ohiorhenuan \&  JD Victor, 
615.8/O01;
S Yu, D Huang, W Singer \& D Nikoli\'c, 
615.14/O07;
MA Sacek et al,
790.1/J12; 
A Tang et al, 
792.4/K27.
\bibitem{hopfield_82}
JJ Hopfield, 
Neural networks and physical systems with emergent collective computational abilities.
{\em Proc Nat'l Acad Sci (USA)} {\bf 79,} 2554--2558 (1982).
\bibitem{tkacik_07a}
G Tka\v{c}ik, {\em Information Flow in Biological Networks} (Dissertation, Princeton University, 2007).
\bibitem{psu_group}
TR Lezon et al,
Using the principle of entropy maximization to infer genetic interaction networks from gene expression patterns.  {\em Proc Nat'l Acad Sci (USA)} {\bf 103,} 19033--19038 (2006).
\bibitem{bialek+ranganathan_07}
W Bialek \& R Ranganathan,  Rediscovering the power of pairwise interactions.  arXiv.org:0712.4397 [q--bio.QM] (2007).
\bibitem{tkacik+al_06}
G Tka\v{c}ik, E Schneidman, MJ Berry II \& W Bialek,  Ising models for networks of real neurons.   q--bio.NC/0611072 (2006).
\bibitem{schneidman+al_03}
E Schneidman, S Still, MJ Berry II \& W Bialek,   
Network information and connected correlations.
{\em Phys Rev Lett} {\bf 91,} 238701 (2003); physics/0307072.
\bibitem{berger+al_96}
AL Berger, S Della Pietra \& VJ Della Pietra,
A maximum entropy approach to natural language processing.
{\em Comp Linguistics} {\bf 22,} 39--71 (1996).
\bibitem{chomsky_56}
N Chomsky,  
Three models for the description of language, {\em IRE Trans Inf Theory}
{\bf IT--2,} 113--124 (1956).
\bibitem{abney_96}
``In one's introductory linguistics course, one learns that Chomsky
disabused the field once and for all of the notion that there was
anything of interest to statistical models of language.  But one usually
comes away a little fuzzy on the question of what, precisely, he
proved.''  S Abney,
Statistical methods and linguistics, in {\em The Balancing Act: Combining
Statistical and Symbolic Approaches to Language,} JL Klavans \& P
Resnik, eds, pp 1--26 (MIT Press, Cambridge, 1996).
\bibitem{pereira_00}
F. Pereira,
Formal grammar and information theory: Together again?
{\em Phil Trans R Soc Lond} {\bf 358,} 1239--1253 (2000).
\bibitem{austen}
The Austen word corpus was created via Project Guttenberg ({\tt  http://www.gutenberg.org}),  combining text from  {\em Emma,  Lady Susan, Love and Friendship, Mansfield Park, Northhanger Abbey,  Persuassion, Pride and Prejudice}  and {\em Sense and Sensibility}.
\bibitem{anc}
{\tt http://www.americannationalcorpus.org}
\bibitem{dictionary}
http://wordlist.sourceforge.net/12dicts-readme.html
\bibitem{technical}
For technical points about finite sample sizes and  error bars, see  N Slonim, GS Atwal, G Tka\v{c}ik \& W Bialek,  Estimating mutual information and multi--information in large networks.Ê arXiv:cs.IT/0502017 (2005).
\bibitem{broderick+al_07}
T Broderick et al,
Faster solutions of the inverse pairwise Ising problem. q-bio.QM/0712.2437 (2007).
\bibitem{darroch+ratcliff_72}
JN Darroch \& D Ratcliff, Generalized iterative scaling for log--linear models.
{\em Ann Math Stat} {\bf 43,} 1470--1480 (1972).
\bibitem{zipf_32}
GK Zipf, {\it Selected Studies of the Principle of Relative Frequency in Language}  (Harvard University Press, Cambridge, 1932).
\bibitem{li_92}
W Li, Random texts exhibit Zipf's--law--like word frequency distribution.  {\em IEEE Trans Inf Theory} {\bf 38,} 1842--1845 (1992).
\bibitem{dehaene+al_05}
S Dehaene, L Cohen, M Sigman \& F Vinckier, The neural code for written words: A proposal. {\em Trends  Cog Sci} {\bf 9,} 335--341 (2005).
\end{thebibliography}
\end{document}